\documentclass[aps,10pt,superscriptaddress,pre,twocolumn,longbibliography]{revtex4-1}
\usepackage{epsfig,graphics,amssymb,amsmath,subeqnarray,color}
\usepackage[applemac]{inputenc}

\usepackage{textcase}
\usepackage{titlesec}
\makeatletter
\newcommand*{\balancecolsandclearpage}{%
  \close@column@grid
  \clearpage
  \twocolumngrid
}
\makeatother
\usepackage[final]{pdfpages}

\begin{document}
\title{Emergent vortices in populations of colloidal rollers}
\author{Antoine Bricard}
\affiliation{Laboratoire de Physique de l'Ecole Normale Sup\'erieure de Lyon, Universit\'e de Lyon and CNRS, 46, all\'ee d'Italie, F-69007 Lyon, France.}
\author{Jean-Baptiste-Caussin}
\affiliation{Laboratoire de Physique de l'Ecole Normale Sup\'erieure de Lyon, Universit\'e de Lyon and CNRS, 46, all\'ee d'Italie, F-69007 Lyon, France.}
\author{Debasish Das}
\affiliation{Department of Mechanical and Aerospace Engineering, University of California San Diego, 9500 Gilman Drive, La Jolla CA 92093-0411, USA}
\author{Charles Savoie}
\affiliation{Laboratoire de Physique de l'Ecole Normale Sup\'erieure de Lyon, Universit\'e de Lyon and CNRS, 46, all\'ee d'Italie, F-69007 Lyon, France.}
\author{Vijayakumar Chikkadi}
\affiliation{Laboratoire de Physique de l'Ecole Normale Sup\'erieure de Lyon, Universit\'e de Lyon and CNRS, 46, all\'ee d'Italie, F-69007 Lyon, France.}
\author{Kyohei Shitara}
\affiliation{Department of Physics,  Kyushu University 33, Fukuoka 812-8581, Japan}
\author{Oleksandr Chepizhko}
\affiliation{Odessa National University, Department for Theoretical Physics, Dvoryanskaya 2, 65026 Odessa, Ukraine}
\affiliation{Universit\'e Nice Sophia Antipolis, Laboratoire J.A. Dieudonn\'e,
UMR 7351 CNRS, Parc Valrose, F-06108 Nice Cedex 02, France} 
\author{Fernando Peruani}
\affiliation{Universit\'e Nice Sophia Antipolis, Laboratoire J.A. Dieudonn\'e,
UMR 7351 CNRS, Parc Valrose, F-06108 Nice Cedex 02, France}
\author{David Saintillan}
\affiliation{Department of Mechanical and Aerospace Engineering, University of California San Diego, 9500 Gilman Drive, La Jolla CA 92093-0411, USA}
\author{Denis Bartolo}
\affiliation{Laboratoire de Physique de l'Ecole Normale Sup\'erieure de Lyon, Universit\'e de Lyon and CNRS, 46, all\'ee d'Italie, F-69007 Lyon, France.}
%
\begin{abstract} 
Coherent vortical motion has been reported in a wide variety of  populations including living organisms (bacteria, fishes, human crowds) and synthetic active matter (shaken grains, mixtures of biopolymers), yet a unified description of the formation and structure of this pattern remains lacking. Here we report the self-organization of motile colloids into a macroscopic steadily rotating vortex. Combining physical experiments and numerical simulations, we elucidate this  collective behavior. We demonstrate that the emergent-vortex structure lives on the verge of a phase separation, and single out the very constituents responsible for this state of polar active matter. Building on this observation, we establish a continuum theory and lay out a strong foundation for the description of vortical collective motion in a broad class of motile populations constrained by geometrical boundaries. 
\clearpage
\end{abstract}
\maketitle

\noindent Building upon the pioneering work of  Vicsek et al.\ \cite{vicsek95}, physicists, mathematicians  and biologists have contemplated the self-organization  of living-organism groups into flocks as an emergent process stemming  from simple interaction rules at the individual level~\cite{TonerReview,vicsekreview,marchettireview}. This idea has  been   supported by   
quantitative trajectory analysis  in animal groups~\cite{Giardinareview,Couzinscience,Theraulaz},  together with a vast number of numerical and theoretical models~\cite{vicsekreview,marchettireview}, and more recently by the
observations of  flocking behavior in ensembles of non-living motile particles such as shaken grains, active colloids, and  mixtures of biofilaments and molecular motors~\cite{Deseigne2010,Baush2010,Bricard2013,Kumar2014}.
From  a physicist's perspective, these various systems are considered as different instances of polar active matter, which encompasses any ensemble of motile bodies endowed with local velocity-alignment interactions.
The current paradigm for flocking physics is the following. Active particles are persistent random walkers, which when dilute form a homogeneous isotropic gas. Upon increasing density, collective motion emerges in the form of spatially localized  swarms that may cruise in a sea of randomly moving particles; further increasing density, a  homogeneous polar liquid forms and spontaneously flows along a well-defined direction~\cite{vicsek95,TonerTu,ChatePRL}. This picture is the outcome of experiments, simulations and theories mostly performed in unbounded or periodic domains. 

Beyond this picture, significant attention has been devoted over the last five years to confined active matter~\cite{vicsekreview,leibler,deseigne,goldstein,Lushi,itaicohen,Aranson,Benjacob,Kudrolli,Mahadevan,Kumar2014,Keber2014,Yang2014,Poon2015}. Confined active particles have consistently, yet not systematically, been reported to self-organize into  vortex-like structures. However, unlike for our understanding of flocking, we are still lacking a unified picture to account for the emergence and structure of such vortex patterns. This situation is mostly due to  the extreme diversity in the nature and  symmetries of the interactions between the active particles that have been hitherto considered. Do active vortices exist only in finite-size systems as in the case of bacterial suspensions~\cite{goldstein}, which lose this beautiful order and display intermittent  turbulent dynamics~\cite{GoldsteinPNAS} when unconfined? What are  the  necessary interactions required to observe and/or engineer  bona fide stationary swirling states of active matter?

In this paper, we answer these questions by considering the impact of geometrical boundaries on the collective behavior of motile  particles endowed with velocity-alignment interactions. Combining quantitative experiments on motile colloids, numerical simulations and analytical theory, we elucidate the phase behavior of {\em polar} active matter restrained by geometrical boundaries.  
We use colloidal rollers which, unlike most of the available biological self-propelled bodies, interact via well established dynamical interactions~\cite{Bricard2013}. We first exploit this unique model system to  show that above a critical concentration populations of motile colloids undergo a non-equilibrium phase transition from an isotropic gaseous state to a novel ordered state where the  entire population   self-organizes  into a single  heterogeneous steadily rotating vortex. This self-organization is {\em not} the consequence of the  finite system size. Rather,  this emergent vortex is a genuine state of polar active matter lying on the verge of a macroscopic phase separation. This novel state is the only ordered phase found when unidirectional directed motion is hindered by convex isotropic boundaries. 
We then demonstrate theoretically that a competition between alignment, repulsive interactions and confinement is necessary  to yield large-scale vortical motion in ensembles of motile particles interacting  via alignment interactions, thereby extending the relevance of our findings to a broad class of active materials.\\ 

\section*{RESULTS}
\begin{figure}[b]
\begin{center}
\includegraphics[width=0.7\columnwidth]{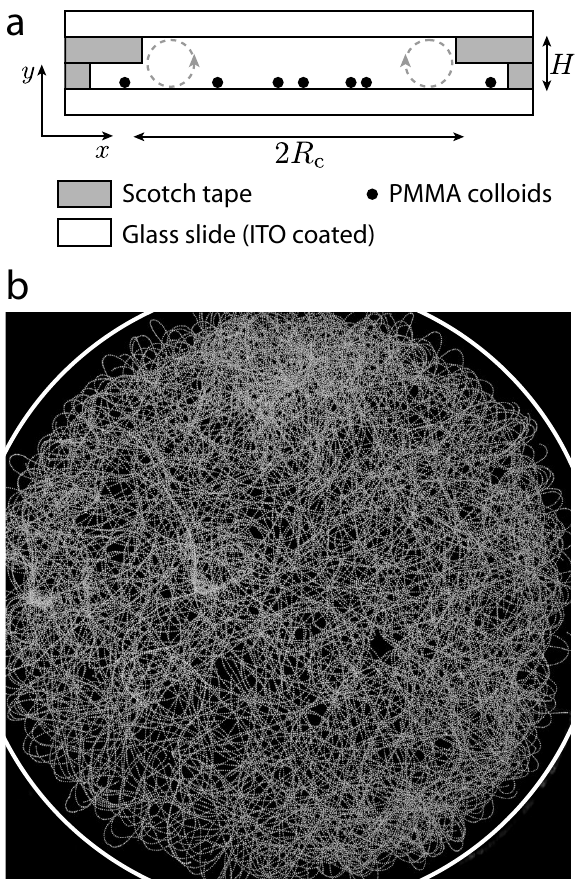}
\caption{{\bf Experimental setup.} ({\bf a}) Sketch of the setup. 5-micron PMMA colloids roll in a microchannel made of two ITO-coated glass slides assembled with double-sided scotch tape. An electrokinetic flow confines the rollers at the center of the device in a circular chamber of radius $R_{\rm c}$. ({\bf b}) Superimposed fluorescence pictures of a dilute ensemble of rollers ($E_0/E_Q=1.1$, $\phi_0=6\times10^{-3}$). The colloids propel only inside a circular disc of radius $R_{\rm c}=1\,\rm mm$ and follow  persistent random walks. \label{fig1}}
\end{center}
\end{figure}
\begin{figure*}
\begin{center}
\includegraphics[width=\textwidth]{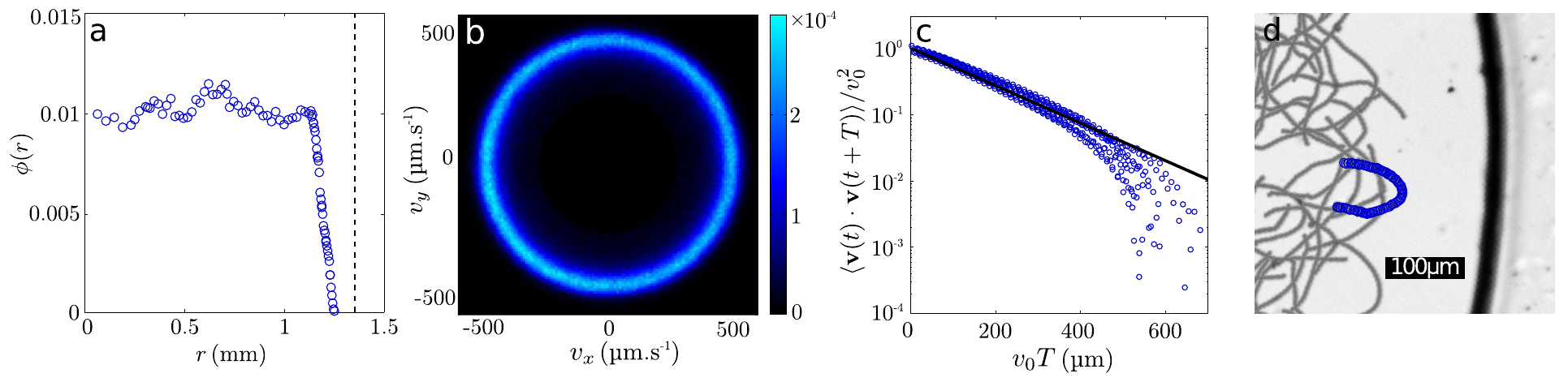}
\caption{{\bf Dynamics of an isolated colloidal roller.} ({\bf a}) Local packing fraction $\phi(r)$, averaged over the azimuthal angle $\phi$, plotted as a function of the radial distance. The dashed line indicates the radius of the circular chamber. ({\bf b}) Probability distribution function of the roller velocities measured from the individual tracking of the trajectories. ({\bf c}) Autocorrelation of the roller velocity $\langle{\bf v}_i(t)\cdot{\bf v}_i(t+T)\rangle$ plotted as a function of $v_0T$ for packing fractions ranging from $\phi_0=6\times10^{-3}$ to $\phi_0=10^{-2}$. Full line: best exponential fit. ({\bf d}) Superimposed trajectories of colloidal rollers bouncing off the edge of the confining circle. Time interval: 5.3 ms. ($E_0/E_Q=1.1$, $\phi_0=6\times 10^{-3}$). Same parameters for the four panels: $R_{\rm c}=1.4\,\rm mm$, $E_0/E_Q=1.1$, $\phi_0=6\times 10^{-3}$.\label{fig2}}
\end{center}
\end{figure*}
\noindent {\bf Experiments. } The experimental setup is fully described in the {\it Methods} section and  in Figs.\ \ref{fig1}a and \ref{fig1}b.  Briefly, we use colloidal rollers powered by the Quincke electrorotation mechanism as thoroughly explained  in~\cite{Bricard2013}. 
 An electric field $\bf E_0$ is applied to insulating colloidal beads immersed in a conducting fluid. Above a critical field amplitude $E_{\rm Q}$,  the symmetry of the electric charge distribution at the bead surface is spontaneously broken. As a result, a net electric torque acts on the beads causing them to rotate at a constant rate around a random axis transverse to the electric field~\cite{Quincke,Taylor,Lemaire}. When the colloids sediment, or  are electrophoretically driven, onto one of the two electrodes, rotation is converted into a net rolling motion along a random direction. Here, we use PMMA spheres of radius  $a=2.4\,\mu\rm m$ immersed in a hexadecane solution.  

 As sketched in Fig.\ \ref{fig1}a, the colloids are handled and observed in a microfluidic device made of double-sided scotch tape and of two ITO-coated glass slides. The ITO layers are  used to apply a uniform DC field ${\bf E}_0=E_0\hat{\bf z}$ in the $z$-direction, with {$E_0=1.6\,\rm V/\mu m$} ($E_0=1.1 E_{\rm Q}$). Importantly, the electric current is nonzero solely  in a disc-shaped chamber at the center of the main channel. As exemplified by the  trajectories shown in Fig.\ \ref{fig1}b and in {\em Supplementary Movie 1},  Quincke rotation is hence restrained to this circular region in which the rollers are trapped.  We henceforth characterize the collective dynamics of the roller population for increasing values of the colloid packing fraction $\phi_0$.\\

\noindent{\bf Individual self-propulsion.}
For area fractions smaller than $\phi^\star=10^{-2}$, the ensemble of rollers uniformly explores the circular confinement as illustrated by the flat profile of the local packing fraction averaged along the azimuthal direction $\phi(r)$ in Fig.~\ref{fig2}a. The rollers undergo uncorrelated persistent random walks as demonstrated in  Figs.\ \ref{fig2}b and~\ref{fig2}c. The  probability distribution of the roller velocities  is isotropic and sharply peaked on the typical speed 
$v_0=493\pm 17\,\rm \mu m.s^{-1}$. In addition, the
 velocity autocorrelation function decays exponentially at short time as expected from a simple model of self-propelled particles having a constant speed  $v_0$ and undergoing rotational diffusion with a rotational diffusivity $D^{-1}=0.31\pm 0.02\,\rm s$ that hardly depends on the area fraction (see {\em Supplementary Discussion 1}).
These quantities correspond to a persistence length of $\ell_{\rm p}=v_0/D=160\,\mu\rm m$ that is about a decade smaller than the confinement radius $R_{\rm c}$ used in our experiments: $0.9\, {\rm mm}<R_{\rm c}< 1.8\, {\rm mm}$. 

At long time, due to the collisions on the disc boundary, the velocity autocorrelation function sharply drops  to 0 as seen in Fig.\ \ref{fig2}c. Unlike swimming cells~\cite{Burke2008,Poon2015}, self-propelled grains~\cite{Kudrolli,Deseigne2010,Mahadevan} or  autophoretic colloids~\cite{Palacci2014}, dilute ensembles of  rollers do not accumulate at the boundary. Instead, they bounce off the walls of this virtual box  as shown in a close-up of a typical roller trajectory in Fig.\ \ref{fig2}d, and  in the {\em Supplementary Movie 1}. As a result, the outer region of the circular chamber is depleted, and the local packing fraction vanishes as $r$ goes to $R_{\rm c}$, Fig.\ \ref{fig2}a. The repulsion from the edges of the circular hole in the microchannel stems from another electrohydrodynamic phenomenon~\cite{shraiman}. When an electric field is applied, a toroidal flow sketched in Fig.\ \ref{fig1}a is osmotically induced by the transport of the electric charges at the surface of the insulating adhesive films. Consequently, a net inward flow  sets in at the vicinity of the bottom electrode.  As the colloidal rollers  are prone to reorient in the direction  of the local fluid velocity~\cite{Bricard2013}, this vortical flow repels the rollers at a distance typically set by the channel height $H$ while leaving unchanged the colloid trajectories in the center of the disc.  This electrokinetic flow will be thoroughly characterized elsewhere.\\
\begin{figure*}
 \vspace*{.05in}
\begin{center}
\includegraphics[width=\textwidth]{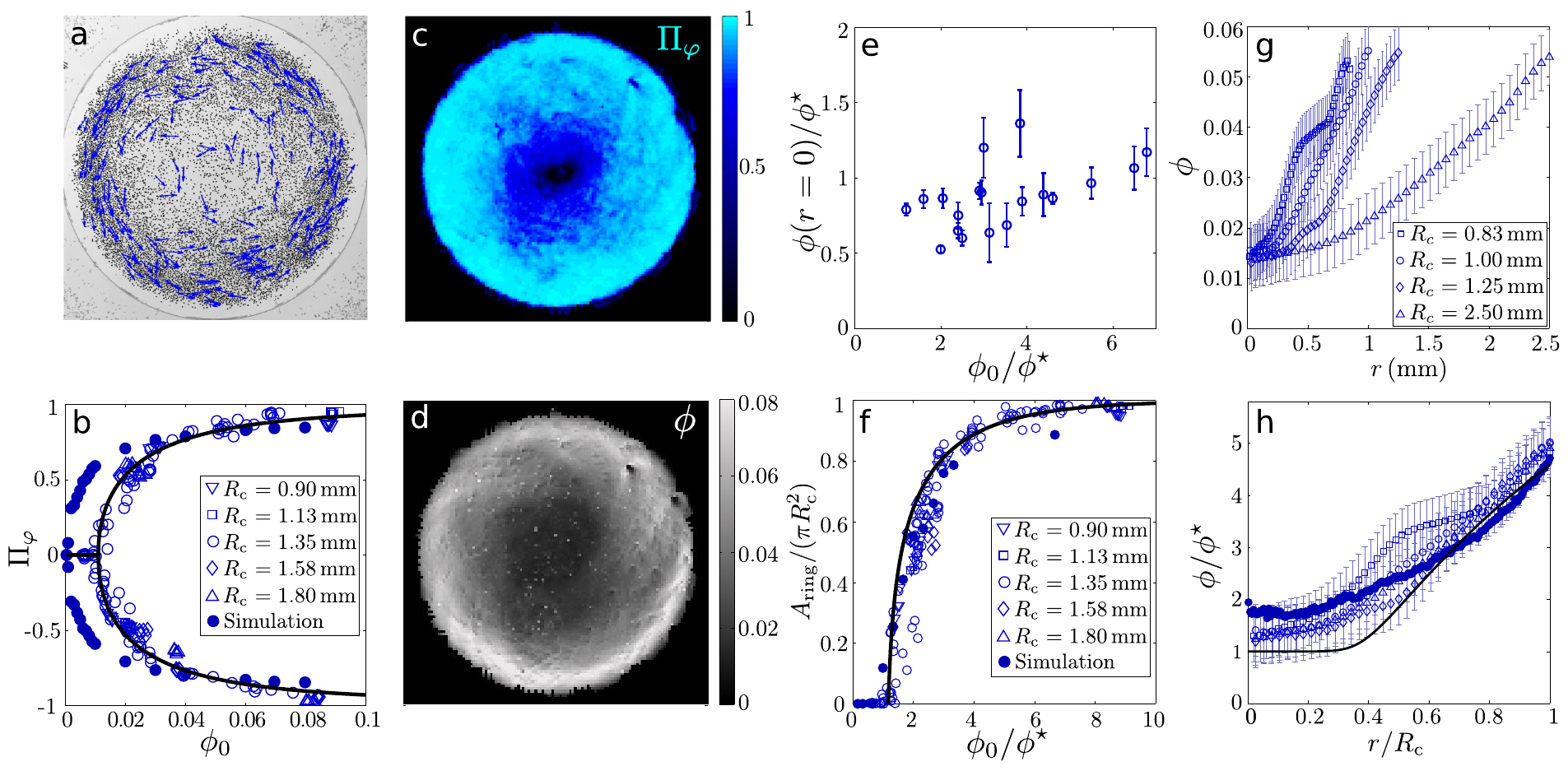}
\caption{{\bf Collective-dynamics experiments.} ({\bf a}) Snapshot of a vortex of rollers. The dark dots show the position of one half of the ensemble of rollers. The blue vectors represent their instantaneous speed ($R_{\rm c}=1.35\,\mathrm{mm}$, $\phi_0=5\times10^{-2}$). ({\bf b}) Average polarization  plotted versus the average packing fraction for different confinement radii. Open symbols: experiments. Full line: best fit from the theory. Filled circles: numerical simulations ($b=3a$, $R_{\rm c}=1\,\rm mm$). ({\bf c}) Time-averaged polarization field ($R_{\rm c}=1.35\,\mathrm{mm}$, $\phi_0=5\times10^{-2}$). ({\bf d}) Time average of the local packing fraction ($R_{\rm c}=1.35\,\mathrm{mm}$, $\phi_0=5\times10^{-2}$). ({\bf e}) Time-averaged packing fraction at the center of the disc, normalized by $\phi^\star$ and plotted versus the average packing fraction. Error bars: one standard deviation.
 ({\bf f}) Fraction of the disc where  $\Pi_\varphi>0.5$ versus the average packing fraction.  Open symbols: experiments. Full line: theoretical prediction with no free fitting parameter. Filled circles: numerical simulations ($b=3a$, $R_{\rm c}=1\,\rm mm$). ({\bf g}) Radial density profiles plotted as a function of the distance to the disc center $r$. All the experiments correspond to $\phi_0=0.032\pm0.002$, error bars: $1\sigma$. ({\bf h}) Open symbols: same data as in ({\bf g}). The radial density profiles are rescaled by $\phi^\star$ and plotted versus the rescaled distance to the center $r/R_{\rm c}$. All the profiles are seen to collapse on a single master curve. Filled symbols: Numerical simulations. Solid line: theoretical prediction.
All the data correspond to $E_0/E_Q=1.1$.\label{fig3}}
\end{center}
\end{figure*}

\noindent{\bf Collective motion in confinement. }
As the area fraction is increased above $\phi^\star$, collective motion emerges spontaneously at the entire population level. When the electric field is applied,  large groups of rollers akin to the band-shaped swarms reported in~\cite{Bricard2013} form and collide. However, unlike what was observed in periodic geometries, the colloidal swarms are merely transient and  ultimately self-organize into a single vortex pattern spanning the entire confining disc as shown in Fig.\ \ref{fig3}a and {\em Supplementary Movie 2}.  
Once formed, the vortex is very robust, rotates steadily and retains an axisymmetric shape. In order to go beyond this qualitative picture, we measured the local colloid velocity field ${\bf v}({\bf r},t)$ and use it to define the polarization field $\mathbf \Pi({\bf r},t)\equiv {\bf v}/v_0$, which quantifies local orientational ordering. The spatial average of $\mathbf \Pi$ vanishes when a coherent vortex forms, therefore we use its projection  $\Pi_\varphi\equiv \langle \mathbf \Pi\cdot\hat{\bf e}_\varphi\rangle_{{\bf r},t}$  along the azimuthal direction as a macroscopic order parameter to probe the transition from an isotropic gas to a polar-vortex state.
As illustrated in Fig.\ \ref{fig3}b,  $\Pi_\varphi(\phi_0)$ displays a sharp bifurcation from an isotropic state with $\Pi_\varphi=0$ to a globally ordered state with equal probability for left- and right-handed  vortices above $\phi_0=\phi^\star$.  Furthermore,  Fig.\ \ref{fig3}b demonstrates that this bifurcation curve does not depend on the confinement radius $R_{\rm c}$. 
The vortex pattern is spatially heterogeneous. The order parameter and density  fields  averaged over time are displayed in Figs.\ \ref{fig3}c and \ref{fig3}d, respectively. At first glance, the system looks phase-separated: a dense and  ordered polar-liquid ring where all the colloids cruise along the azimuthal direction  encloses a  dilute and weakly ordered core at the center of the disc.  We shall also stress that regardless of the average packing fraction, the  packing fraction in the vortex core is measured to be very close to $\phi^\star$, the average concentration below which the population is in a gaseous state, see Fig.\ \ref{fig3}e. 
This phase-separation picture  is  consistent with the variations of the area occupied by the ordered outer ring, $A_{\rm ring}$,  for different confinement radii $R_{\rm c}$, as shown in Fig.\ \ref{fig3}e. We define $A_{\rm ring}$ as the area of the region where the order parameter exceeds 0.5, and none of the results reported below depend on this arbitrary choice for the definition of the outer-ring region.
$A_{\rm ring}$ also bifurcates as $\phi_0$ exceeds $\phi^\star$, and increases with $R_{\rm c}$. Remarkably, all the bifurcation curves collapse on a single master curve when  $A_{\rm ring}$ is rescaled by the overall confinement area $\pi R_{\rm c}^2$, Fig.\ \ref{fig3}f. In other words, the strongly polarized outer ring always occupies the same area fraction irrespective of the system size, as would a molecular liquid coexisting with a vapor phase at equilibrium. However, 
if the system were genuinely phase-separated, one should be able to define an interface between the dense outer ring and the dilute inner core, and this interface should have a constant width. This requirement is not borne out by our measurements. The shape of the radial density profiles of the rollers in Fig.\ \ref{fig3}g indeed makes it difficult to unambiguously define two homogeneous phases separated by a clear interface. Repeating the same experiments in discs of increasing radii, 
 we found  that the density profiles are self-similar, Fig.\ \ref{fig3}h.  The width of the region separating the strongly polarized outer ring from the  inner core scales with the system size, which is the only characteristic scale of the vortex patterns.  The colloidal vortices therefore correspond to a monophasic yet spatially heterogeneous liquid state. 

In order to elucidate the physical mechanisms responsible for this intriguing structure, we now introduce a theoretical model that we solve both numerically and analytically.\\

\noindent {\bf Numerical simulations. }
\begin{figure*}
 \vspace*{.05in}
\begin{center}
\includegraphics[width=\textwidth]{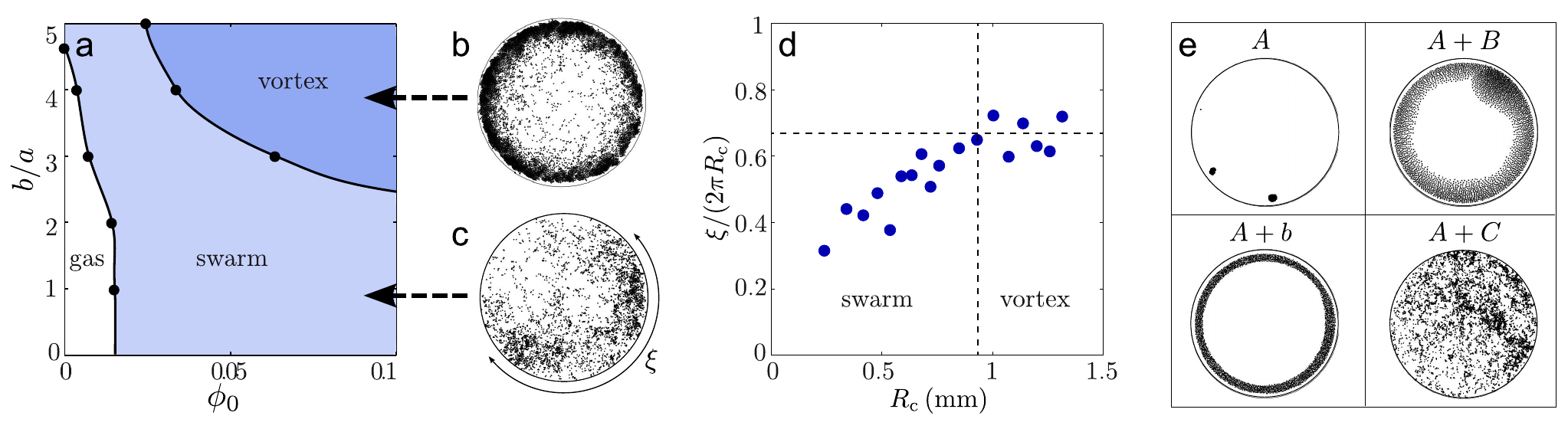}
\caption{{\bf Collective-dynamics  simulations.} ({\bf a}) The numerical phase diagram of the confined population is composed of three regions: isotropic gas (low $\phi_0$, small $b$), swarm coexisting with a gaseous phase (intermediate $\phi_0$ and $b$), and vortex state (high $\phi_0$ and $b$). $R_{\rm c}=0.5\,{\rm mm}$. ({\bf b}) Snapshot of a vortex state. Numerical simulation for $\phi_0=0.1$, and $b=5a$. ({\bf c}) Snapshot of a swarm. Numerical simulation for $\phi_0=4.5\times10^{-2}$, and $b=2a$. ({\bf d}) Variation of the density correlation length as a function of $R_{\rm c}$. Above $R_{\rm c}=1\,\rm mm$, $\xi$ plateaus and a vortex is reached ($\phi_0=3\times 10^{-2}$, $b=3a$). ({\bf e}) Four numerical snapshots of rollers interacting via: alignment interactions only ($A$),  alignment interactions and repulsive torques ($A+B$, where the magnitude of $B$ is 5 times the experimental value),  alignment and excluded volume interactions ($A+b$, where the repulsion distance is $b=5a)$,  alignment and the $C$-term in Eq. 3 ($A+C$). Polarized vortices emerge solely when repulsive couplings exist ($A+B$ and $A+b$). \label{Fig4}}
\end{center}
\end{figure*}
The Quincke rollers are electrically powered and move in a viscous fluid, and hence  interact at a distance  both hydrodynamically and electrostatically. In \cite{Bricard2013}, starting from the Stokes and Maxwell equations, we established the equations of motion of a dilute ensemble of Quincke rollers within a pairwise additive approximation.  When isolated, the $i$th roller  located at ${\bf r}_i$ moves at a speed $v_0$ along  the direction $\hat{\bf p}_i=(\cos \theta_i,\sin \theta_i)$  opposite to the in-plane component of the electrostatic dipole responsible for Quincke rotation~\cite{Bricard2013}. When interacting via contact and electrostatic repulsive forces, the roller velocity and orientation are related by: 
\begin{align}
&\partial_t{\bf r}_i= v_0 \hat{\bf p}_i -\partial_{{\bf r}_i}\sum_{j\neq i}{\cal H}_{\rm rep}({\bf r}_i-{\bf r}_j).
\label{Eq1}
\end{align} 
Inertia is obviously ignored, and for the sake of simplicity we  model all the central forces acting on the colloids as an effective hard-disc exclusion ${\cal H}_{\rm rep}$ of range $b$. In addition, $\theta_i$ follows an overdamped dynamics in an effective angular potential capturing both the electrostatic and hydrodynamic torques acting on the colloids~\cite{Bricard2013}:
\begin{align}
& \partial_t\theta_i(t)=\partial_{\theta_i}\sum_{j\neq i}{{\cal H}({\bf r}_i-{\bf r}_j;\hat {\bf p}_i,\hat {\bf p}_j})+\xi_i.
\label{Eq2}
\end{align} 
The $\xi_i$'s account for rotational diffusion of the rollers. They are uncorrelated white noise variables with zero mean and variance $\langle \xi_i(t) \xi_j(t') \rangle = 2 D \delta(t-t')\delta_{ij}$. The effective potential in Eq.\ \textbf{\ref{Eq2}} is  composed of three terms with clear physical interpretations:
\begin{align}
   \label{Eq3}     \mathcal H ({\bf r},\hat { \bf p}_i,  \hat { \bf p}_j) &= A(r) \, \hat { \bf p}_j \cdot \hat { \bf p}_i+ B(r) \, \hat { \bf r} \cdot \hat { \bf p}_i \\
        &+ C(r) \, \hat { \bf p}_j \cdot (2\hat { \bf r}\hat { \bf r} - \mathbb I  )\cdot \hat { \bf p}_i,\nonumber
\end{align}
where $\hat{\mathbf{r}}=\mathbf{r}/r$. The symmetry of these interactions is not specific to colloidal rollers and could have been anticipated phenomenologically exploiting both the translational invariance and the polar symmetry of the surface-charge distribution of the colloids~\cite{Caussin20142}.
The first term promotes alignment and is such that the effective potential is minimized when interacting  rollers propel along the same direction. $A(r)$ is positive, decays exponentially  with $r/H$, and  results both from hydrodynamic and electrostatic interactions. The second term  gives rise to repulsive {\em torques}, and is minimized when  the  roller  orientation points away from its interacting neighbor. $B(r)$ also decays exponentially with $r/H$  but solely stems from  electrostatics.  The third term has a  less intuitive meaning, and promotes the alignment of $\hat{\bf p}_i$ along a dipolar field oriented along  $\hat{\bf p}_j$. This term is a combination of hydrodynamic and electrostatic interactions, and includes a  long-ranged contribution. 

The functions $A(r)$, $B(r)$, and $C(r)$ are provided in the {\em Supplementary Discussion 2}. As it turns out, all the physical parameters (roller velocity, field amplitude, fluid viscosity, etc.)  that are needed to compute their  exact expressions  have been measured, or estimated up to  logarithmic corrections, see {\em Supplementary Discussion 2}. We are then left with a model having a single free parameter that is the range, $b$, of the repulsive {\em forces} between colloids. 
We numerically solved this model in  circular simulation boxes of radius $R_{\rm c}$ with reflecting boundary conditions using an explicit Euler scheme with adaptive time-stepping. All the numerical results are discussed using the same units as in the experiments to facilitate quantitative comparisons. 

The simulations revealed a richer phenomenology than the experiments, as captured by the phase diagram in Fig.~\ref{Fig4}a corresponding to $R_{\rm c}=0.5\,\rm mm$. By systematically varying the range of the repulsive forces and the particle concentration, we found that the $(\phi_0,b)$ plane  is typically divided into three regions.  At small packing fractions, the particles hardly interact and form an isotropic gaseous phase. At high fractions,  after transient dynamics strikingly similar to that observed in the experiments, the rollers self-organize into a macroscopic vortex pattern, Fig.\ \ref{Fig4}b  and {\em Supplementary Movie 3}. However, at intermediate densities, we found that collective motion emerges in the form of a macroscopic swarm cruising around the circular box through an ensemble of randomly moving particles, Fig.\ \ref{Fig4}c and {\em Supplementary Movie 4}. These swarms are akin to the  band patterns consistently reported for polar active particles at the onset of collective motion in periodic domains~\cite{ChatePRL,Bricard2013}.  This seeming conflict between our experimental and numerical findings is solved by looking at the variations of the swarm length $\xi_{\rm s}$ with the confinement radius $R_{\rm c}$ in Fig.\ \ref{Fig4}d.  We define $\xi_{\rm s}$ as  the correlation length of the density fluctuations in the azimuthal direction.  The angular  extension of the swarms $\xi_{\rm s}/R_{\rm c}$ increases linearly with the box radius. Therefore, for a given value of the interaction parameters, there exists a critical box size above which the population undergoes a direct transition from a gaseous to an axisymmetric vortex state. For $b=3a$, which  was measured to be the typical interparticle distance in the polar liquid state~\cite{Bricard2013},  this critical confinement is $R_{\rm c}=1\,\rm mm$. This value is close to the smallest radius accessible in our experiments where localized swarms were never observed, thereby solving the apparent discrepancy with the experimental phenomenology.  

More quantitatively, we  systematically  compare our numerical and  experimental measurements in Figs.\ \ref{fig3}b and \ref{fig3}c for $R_{\rm c}=1\,\rm mm$. Even though a number of simplifications were needed to establish Eqs.\ \textbf{\ref{Eq1}},~\textbf{\ref{Eq2}} and~\textbf{\ref{Eq3}}~\cite{Bricard2013}, the simulations  account very well for the sharp bifurcation yielding the vortex patterns as well as their self-similar structure.  This last point is  proven quantitatively in Fig.\ \ref{fig3}h, which demonstrates that the concentration increases away from the vortex core, where $\phi(r=0)=\phi^{\star}$, over a scale that is solely set by the confinement radius.  We shall note however that the numerical simulations underestimate the critical packing fraction $\phi^\star$ at which collective motion occurs, which is not really surprising given the number of approximations required to establish the interaction parameters in the equations of motion Eq.~3. 
 We  unambiguously conclude from this set of results that Eqs.\ \textbf{1}, \textbf{2} and \textbf{3} include all the physical ingredients that chiefly dictate the collective dynamics of the colloidal rollers. We  now exploit the opportunity offered by the numerics to  turn on and off the four roller-roller interactions one at a time, namely the alignment torque, $A$, the repulsion torque $B$ and force $b$, and the dipolar coupling $C$. Snapshots of the resulting particle distributions are reported in Fig.\ \ref{Fig4}e. None of these four interactions alone yields  a coherent  macroscopic vortex.  We stress  that  when the particles solely interact via pairwise-additive alignment torques, $B=C=b=0$, the population condenses into a single compact polarized swarm. Potential velocity-alignment interactions are {\em not} sufficient to yield macroscopic vortical motion. We  evidence in Fig.\ \ref{Fig4}e (top-right and bottom-left panels) that  the combination of alignment ($A\neq0$) and of repulsive interactions ($B\neq 0$ and/or $b\neq0$) is necessary and sufficient to observe spontaneously flowing vortices. \\
 
 \noindent{\bf Analytical theory. }
Having identified the very ingredients necessary to account for our observations, we can now gain more detailed physical insight into the spatial structure of the vortices by constructing a minimal hydrodynamic theory. We start from Eqs.\ \textbf{1}, \textbf{2} and \textbf{3}, ignoring the $C$ term in Eq.\ \textbf{3}. The model can be further simplified by inspecting the experimental variations of the individual roller velocity with the local packing fraction, see {\em Supplementary Figure 1}. The roller speed only displays  variations of 10$\%$ as $\phi(\bf r)$ increases from $10^{-2}$ to $4\times10^{-2}$. These minute variations suggest ignoring the contributions of the repulsive forces in Eq.\ \textbf{1}, and solely considering  the interplay between the alignment and repulsion torques on the orientational dynamics of Eq.\ \textbf{2}.  
These simplified equations of motion are coarse-grained following  a conventional  kinetic-theory framework   reviewed in~\cite{marchettireview} to establish the equivalent to the Navier-Stokes equations for this  two-dimensional active fluid. 
In a nutshell, the  two observables we need to describe are the local area fraction $\phi$ and the local momentum field  $\phi \mathbf\Pi$. They are related to the first two angular moments of the one-particle distribution function $\psi({\bf r},\hat{\bf p},t)=\pi a^2\langle\sum_i \delta({\bf r}-{\bf r}_i)\delta(\hat{\bf p}-\hat{\bf p}_i)\rangle$, which evolves according to a  Fokker-Plank equation derived from the conservation of $\psi$ and Eqs.\ \textbf{\ref{Eq1}} and \textbf{\ref{Eq2}}. This  equation is then recast into an infinite hierarchy of equations for the angular moments of $\psi$. The two first equations of this hierarchy, corresponding to the mass conservation equation and to the momentum dynamics, are akin to the continuous theory first introduced phenomenologically by Toner and Tu~\cite{marchettireview,TonerReview}:
\begin{align}
\partial_t\phi+v_0\nabla\cdot \left(\phi \mathbf \Pi \right )&=0,\label{Eq4}\\
\partial_t\left(\phi \mathbf \Pi \right )+v_0\nabla\cdot\left(\phi \mathbf Q+\frac{\phi}{2}\mathbb I\right)&={\mathbf F}(\phi\,\mathbf, \mathbf\Pi,\mathbf Q),\label{Eq5}
\end{align}
where ${ \mathbf Q}$ is the usual nematic order parameter.  The meaning of the first equation is straightforward, while the second calls for some clarifications. The divergence term on the left-hand side of Eq.\ \textbf{\ref{Eq5}} is a  convective kinematic term associated with the self-propulsion of the particles. The force field $\mathbf F$ on the right-hand side would vanish  for non-interacting particles. Here,  at first order in a gradient expansion, $\mathbf F$ is given by:
\begin{align}
{\mathbf F}=&- D \, \phi \mathbf \Pi +\alpha \, \phi^2(\mathbb I - 2 \mathbf Q)\cdot \mathbf \Pi - \beta\phi (\mathbb I - 2 \mathbf Q) \cdot \nabla \phi. \label{Eq6}
\end{align}
This force field  has a clear physical interpretation. The first term reflects the damping of the polarization by the rotational diffusion of the rollers. The second term,  defined by the time rate $\alpha = (\int_{r > 2a}  rA(r) {\rm d} r)/a^2$, echoes the alignment rule at the microscopic level and promotes a nonzero local polarization. The third term,  involving $\beta =(\int_{r > 2a}  r^2B(r) {\rm d} r)/(2 a^2)$, is an anisotropic pressure reflecting the repulsive interactions between rollers at the microscopic level. 
Eqs.\ \textbf{\ref{Eq4}} and \textbf{\ref{Eq5}} are usually complemented by a dynamical equation for $\mathbf Q$ and a closure relation. This additional approximation, however, is not needed to demonstrate the existence of vortex patterns and to rationalize their spatial structure. 

Looking for axisymmetric steady  states, it readily follows from mass conservation, Eq.\ \textbf{\ref{Eq4}},  that the local fields must take  the simple forms: $\phi = \phi(r)$, $\mathbf \Pi = \Pi_\varphi(r) \mathbf e_\varphi$ and $\mathbf Q = Q(r)(\mathbf e_\varphi \mathbf e_\varphi - \mathbf e_r \mathbf e_r)$, where $Q(r)>0$.  We also infer the relation $ \left[ -D+ \alpha\phi(1-2Q) \right]\phi \Pi_\varphi=0$ from the  projection of the momentum equation, Eq.\ \textbf{\ref{Eq5}}, on the azimuthal direction. This relation tells us that the competition between rotational diffusion and local alignment results in a mean-field transition from an isotropic state with $\Pi_\varphi=0$ to a polarized vortex state with $\Pi_\varphi\neq0$ and $Q=\frac{1}{2}\left(1-D/(\alpha\phi)\right)$. This transition occurs when $\phi$ exceeds $\phi^\star\equiv D/\alpha$,  the ratio of the rotational diffusivity to the alignment strength at the hydrodynamic level. In addition, the  projection of Eq.\ \textbf{\ref{Eq5}} on the radial direction sets the spatial structure of the ordered phase:
\begin{equation}
\label{Eq7}
2\frac{v_0}{r}Q -{\beta} \left( 1+2Q \right) \frac{{\rm d} \phi}{{\rm d}r}=0,
\end{equation}
with again $Q=\frac{1}{2}\left(1-\phi^\star/\phi\right)$ in the ordered polar phase.
 This equation has  a clear physical meaning and expresses the balance between the centrifugal force arising from the advection of momentum along a circular trajectory  and the anisotropic pressure induced by the repulsive interactions between rollers. It has an implicit solution given by   
\begin{equation}
\frac{r}{r^\star} = \exp\left[ 2 \Lambda \frac{\phi}{\phi^\star} + \Lambda \log\left( \frac{\phi}{\phi^\star} - 1 \right) \right].
\label{Eq8}
\end{equation}
$\phi(r)$ is therefore parametrized by the dimensionless number $\Lambda\equiv \phi^\star \frac{\beta}{v_0}$ reflecting the interplay between self-propulsion and repulsive interactions. Given the experimental values of the microscopic parameters, $\Lambda$ is much smaller that unity ($\Lambda\sim 0.08$). 
An asymptotic analysis reveals that $r^\star$ is the typical core radius of the vortex. For $r<r^\star$, the density increases  slowly as   $\phi \sim \phi^\star \left[ 1 + (r/r^\star)^{1/\Lambda} \right]$ for all $\phi_0$ and  $R_{\rm c}$. As $r$ reaches $r^\star$, it increases significantly and then growths logarithmically as $\phi \sim v_0/\beta \log(r/r^\star)$ away from the vortex core. However,  $r^\star$ is an integration constant which is solely defined via the mass conservation relation:
$
\pi R^2_{\rm c}\phi_0 =  \int_0^{R_{\rm c}} 2\pi r\phi(r)  \, {\rm d} r $ and therefore only depends on $\phi_0$ and $R_{\rm c}$. $r^\star$ does not provide any intrinsic structural scale, and the vortex patterns formed in different confinements are predicted to be self-similar in agreement with our experiments and  simulations despite the simplification made in the model, Fig.\ \ref{fig3}e. In addition, Eq.\ \textbf{\ref{Eq8}} implies that the rollers self-organize by reducing their  density at the center of the vortex  down to $\phi=\phi^\star$,  the mean area fraction at the onset of collective motion, again in excellent agreement with our measurements in Fig.\ \ref{fig3}e. 

In order to characterize the orientational structure of the vortices,  an additional closure relation is now required. The simplest possible choice consists in neglecting correlations of the orientational fluctuations, and therefore assuming  $Q=\Pi_\varphi^2/2$. This choice implies that 
\begin{equation}
\Pi_\varphi(r)=\sqrt{1-\phi^\star/\phi(r)}.\label{Eq9}
\end{equation} 
Eqs.\ \textbf{\ref{Eq8}} and \textbf{\ref{Eq9}} provide a very nice fit of the experimental polarization curve as shown in Fig.\ \ref{fig3}b, and therefore capture both the pitchfork bifurcation scenario at the onset of collective motion and the saturation of the polarization at high packing fractions.  The best fit is obtained for values of $\phi^\star$ and $\beta$ respectively five and two  times larger than those deduced from the microscopic parameters. Given the number of simplifications needed to establish both the microscopic and hydrodynamic models, the agreement is very convincing. We are then left with a hydrodynamic theory with no free fitting parameter, which we use to compute the  area fraction of the outer polarized ring  where $\Pi_\varphi>0.5$. The comparison with the experimental data in Fig.\ \ref{fig3}f is excellent.

Furthermore, Eqs.\ \textbf{\ref{Eq8}} and \textbf{\ref{Eq9}} predict that the rollers are on the verge of a phase separation. If the roller fraction in the vortex core were smaller ($\phi<\phi^\star$), orientational order could not be supported and an isotropic bubble would nucleate in a polar liquid. This phase separation is avoided by the self-regulation of $\phi(r=0)$ at $\phi^\star$. 
 
\section*{DISCUSSION}
Altogether our theoretical results  confirm that the vortex patterns stem  from the interplay between self-propulsion, alignment, repulsion and confinement. Self-propulsion and alignment interactions promote a  global azimuthal flow. The repulsive interactions prevent condensation of the population on the geometrical boundary and allow for  extended vortex patterns. If the rollers were not confined, the population would evaporate as self-propulsion induces a centrifugal force despite the absence of inertia.

We close this discussion by stressing on the generality of this scenario.  { Firstly, the vortex patterns do not rely on the perfect rotational symmetry of the boundaries. As illustrated in  {\em Supplementary Figure 2 } the same spatial organization is observed for a variety of convex polygonal geometries. However, strongly anisotropic, and/or strongly non-convex confinements can yield other self-organized states such as vortex arrays which we will characterize elsewhere. 
Secondly, neither the nature of the repulsive couplings nor the symmetry of the interactions yielding collective motion are crucial,} thereby making the above results relevant to a much broader class of experimental systems. For instance, self-propelled  particles endowed with  nematic alignment rules are expected to display the same large-scale phenomenology. The existence of a centrifugal force  does not rely on the direction of the  individual trajectories. Shaken rods,  concentrated suspensions of bacteria, or motile biofilaments, among other possible realizations, are expected to have a similar phase behavior. 
Quantitative local analysis of their spatial patterns~\cite{Baush2010,Kumar2014,leibler,deseigne,goldstein}  would make it possible to further test and elaborate our understanding of the structure of confined active matter. { For instance, the polar order found in confined bacteria is destroyed upon increasing the size of the confinement. The analysis of the spacial distribution of the bacteria could be used to gain insight on the symmetries and the magnitude of the additional interactions mediated  by the host fluid which are responsible for  the emergence of bacterial turbulence~\cite{goldstein}.}

In conclusion, we  take advantage of a model experimental system where ensembles of self-propelled colloids with well-established interactions self-organize into macrosopic vortices when confined by circular geometric boundaries. We  identify the physical mechanism that chiefly dictates this emergent behavior. Thanks to a combination of numerical simulations and analytical theory, we  demonstrate that orientational couplings alone cannot account for collective circular motion. Repulsion between the motile individuals is necessary to balance the centrifugal flow intrinsic to any ordered active fluid and to stabilize heterogeneous yet monophasic states in a broad class of active fluids. { A natural challenge is to extend this description to the compact vortices   observed in the wild, e.\ g. in shoals of fish. In the absence of confining boundaries, the centrifugal force has to be balanced by additional density-regulation mechanisms~\cite{Couzinattractionrepulsion,Turner}. A structural investigation akin to the one introduced here for roller vortices could be a powerful tool to shed light on  density regulation  in natural flocks which remains to be elucidated. }

\section*{METHODS}
\noindent {\bf Experiments.}
We use fluorescent PMMA colloids (Thermo scientific G0500, 2.4$\,\mu$m radius), dispersed in a 0.15 mol.L$^{-1}$ AOT/hexadecane solution. The suspension is injected in a wide microfluidic chamber made of double-sided scotch tapes. The tape is sandwiched between two ITO-coated glass slides (Solems, ITOSOL30, 80 nm thick). 
An additional layer of scotch tape including a hole having the desired confinement geometry  is added to the upper ITO-coated slide.  The holes are made with a precision plotting cutter (Graphtec robo CE 6000). The gap between the two ITO electrodes is constant over the entire chamber $H=220\,\mu \rm m$.
 The electric field is applied by means of a voltage amplifier (Stanford Research Systems, PS350/5000V-25W).  All the measurements were performed 5 minutes after the beginning of the rolling motion, when a steady state was reached for all the observables. 
 
 The colloids are observed with a 4X microscope objective for  particle tracking, PIV and number-density measurements. High-speed 
 movies are recorded with a CMOS camera (Basler ACE) at a frame rate of 190fps. All  images are 2000$\times$2000 8-bit pictures. 
 The particles are detected to sub-pixel accuracy and the particle trajectories are reconstructed using a MATLAB version of a conventional  tracking code \cite{Grier}. The PIV analysis was performed with the mpiv MATLAB code. A block size of $44\,\mu$m was used.
 
\vspace{0.2cm}\noindent {\bf Numerical simulations.} The simulations are performed by numerically integrating the equations of motion Eqs.~\textbf{1} and~\textbf{2}. Particle positions and rolling directions are initialized randomly inside a circular domain. Integration is done using an Euler scheme with an adaptive time step $\delta t$, and the diffusive term in the equation for the rotational dynamics is modeled as a Gaussian variable with zero mean and with variance $2D/\delta t$. Steric exclusion between particles is captured by correcting particle positions after each time step so as to prevent overlaps. Bouncing off of particles at the confining boundary is captured using a phenomenological torque that reorients the particles towards the center of the disc; the form of the torque was chosen so at the reproduce the bouncing trajectories observed in the experiments.

\vspace{0.3cm}
\noindent {\bf Acknowledgments}
We benefited from valuable discussions with Hugues Chat\'e, Nicolas Desreumaux, Olivier Dauchot, Cristina Marchetti, Julien Tailleur and John Toner. This work was partly funded by the ANR program MiTra, and Institut Universitaire de France. D.S. acknowledges partial support from the Donors of the American Chemical Society Petroleum Research Fund and from NSF CAREER Grant No. CBET-1151590. K.S. was supported by the JSPS Core-to-Core Program ``Non-equilibrium dynamics of soft matter and information''. 

\vspace{0.3cm}
\noindent{\bf Author contributions}
A.B. and V.C. carried out the experiments and processed the data. 
D.D., C.S., O.C., F.P., and  D.S. carried out the the numerical simulations.
J.-B.C., K.S. and D.B. established the analytical model.
All the authors discussed and interpreted results. D.B., J.-B.C. and D.S.  wrote the manuscript. D.B. conceived the project.
A.B. and J.-B.C. have equally contributed to this work.

\vspace{0.3cm}
\noindent{\bf Additional information}
Supplementary Information accompanies this paper.

\vspace{0.3cm}
\noindent{\bf Competing financial interests:} The authors declare no competing financial interests.

\balancecolsandclearpage

\onecolumngrid
\appendix

\titleformat{\section}{\center \small \bfseries}{\thesection. }{0pt}{\MakeTextUppercase}   \titlespacing*{\section}{0pt}{6ex plus 1ex minus .2ex}{4ex plus .2ex}
\titleformat{\subsection}{\center \small \bfseries}{\thesubsection. }{0pt}{}    \titlespacing*{\subsection} {0pt}{3.25ex plus 1ex minus .2ex}{3ex plus .2ex}

\begin{center}
{\bf \Large Supplementary Information}
\end{center}
\vspace{0.1cm}


\noindent{\bf \large Supplementary Figures}

\begin{figure}[h!]
\begin{center}
\includegraphics[width=0.34\columnwidth]{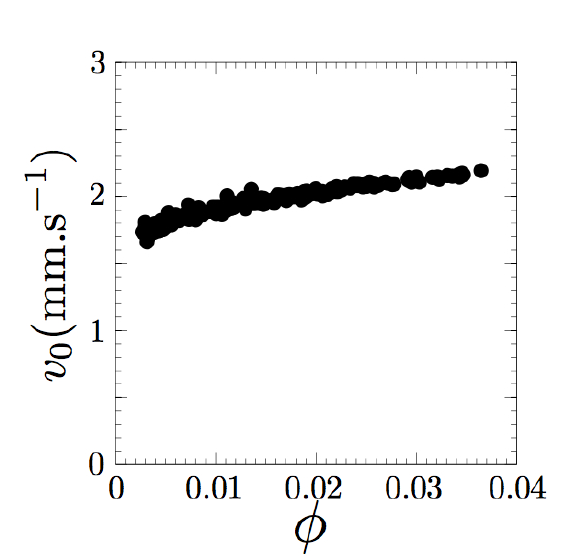}
\caption{Variation of the roller velocity with the local area fraction. $E/E_{\rm Q}=1.4$. As $\phi$ varies from $10^{-2}$ to $4 \times 10^{-2}$, $v_0$ only increases by $\sim 10\%$.}
\label{figS2}
\end{center}
\end{figure}

\begin{figure}[h!]
\begin{center}
\includegraphics[width=0.6\columnwidth]{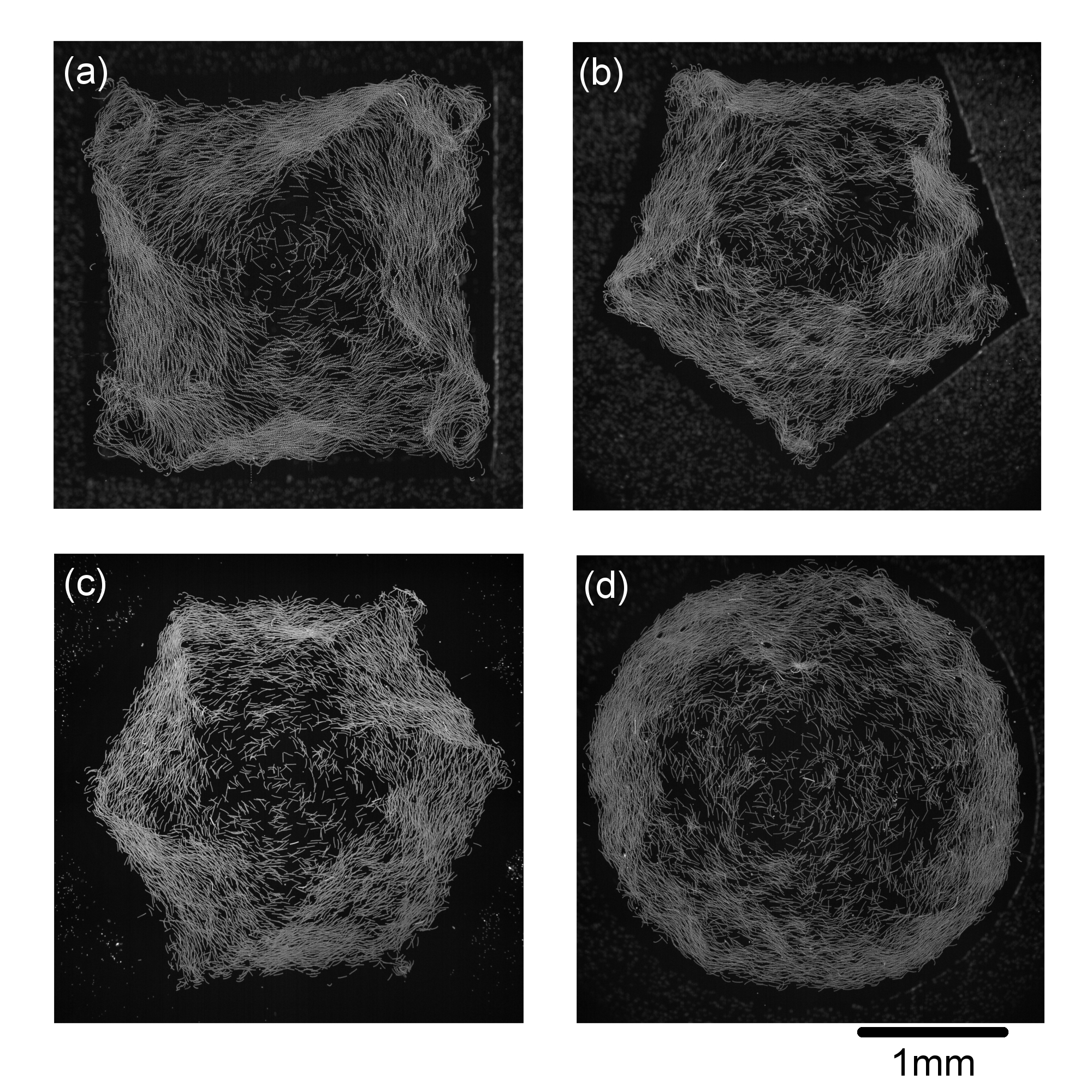}
\caption{Superimposed pictures of colloidal rollers forming a vortex patterns in polygonal geometries. Packing fraction:  $\phi_0=3\times 10^{-2}$. $E/E_{\rm Q}=1.4$. The self organization of the population  is robust to the shape of the confinement. These pictures show that the same heterogeneous vortex pattern emerge in polygonal confinements which do not have a perfect rotational symmetry.  The center of the vortex is dilute and less ordered that the outer region where polar order is pronounced. The shape of the polygonal geometries is reflected by the velocity and density profiles yet the salient features of the vortex is preserved.}
\label{figS2}
\end{center}
\end{figure}

\begin{figure}[!h]
 \vspace*{.05in}
\begin{center}
\includegraphics[width=0.4\columnwidth]{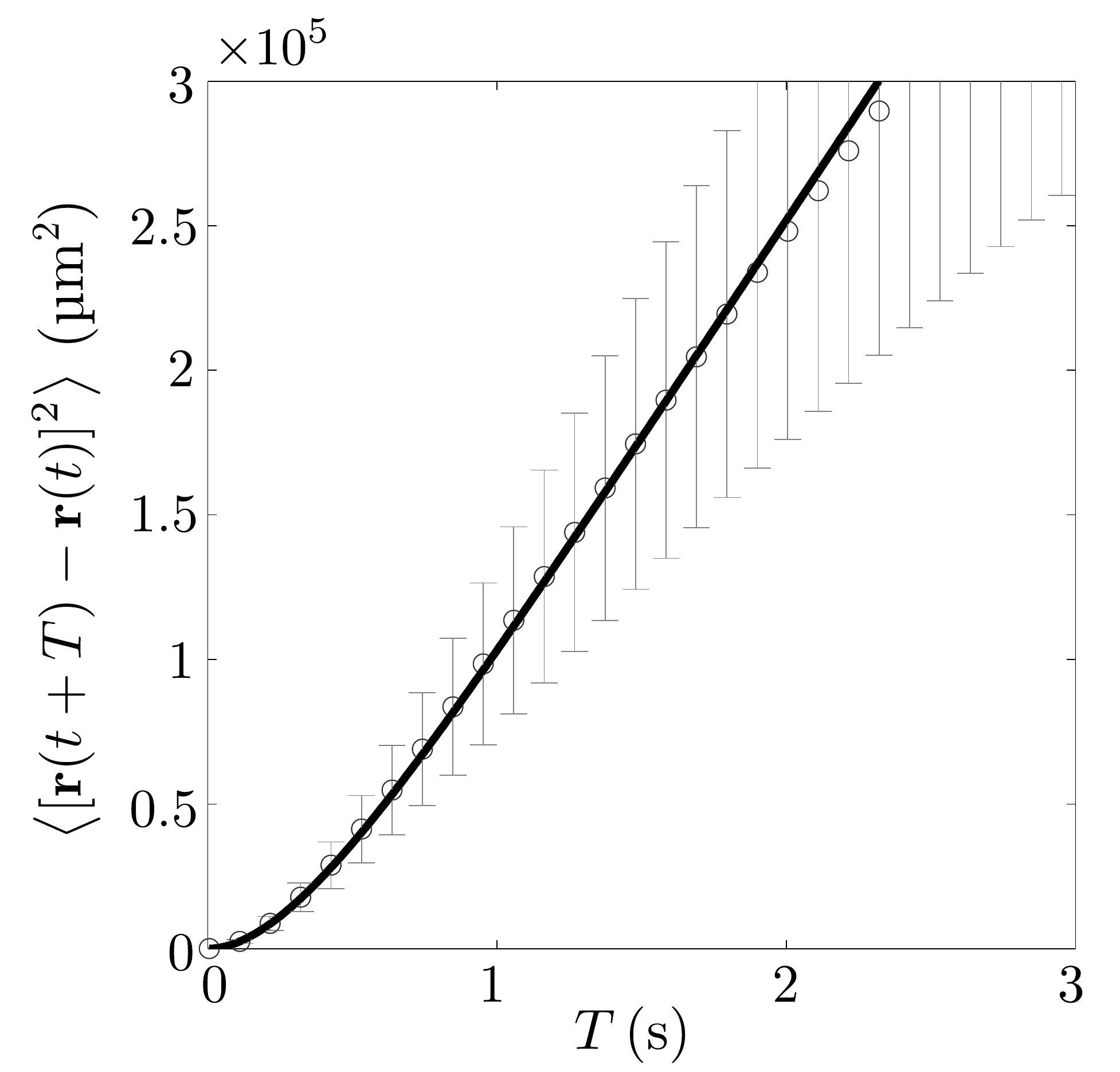}
\caption{Mean-squared displacement of the rollers, $\langle \left[ \mathbf r_i(t+T) - \mathbf r_i(t) \right]^2 \rangle_{i,t}$ plotted as a function of the lag time $T$. Open symbol: experiments (error bars 1SD). Full line: theoretical prediction.}
\label{figS1}
\end{center}
\end{figure}

\pagebreak


\noindent{\bf \large Supplementary Discussions}

\vspace{0.4cm}

\noindent{\bf Supplementary Discussion 1 $\vert$ Dynamics of a single roller}\\

At low packing fraction, $\phi_0 \ll \phi^\star$, the colloids behave as non-interacting persistent random walkers. Their motion is described by:
 \begin{align}
 	\partial_t \mathbf r_i &= v_0 \hat{\mathbf p}_i = v_0 (\cos \theta_i, \sin \theta_i) ,\\
	\partial_t \theta_i &= \xi_i(t) ,
 \end{align}
 where $\xi_i$ is a white noise with zero mean and variance $\langle \xi_i(t) \xi_i(t') \rangle = 2 D \delta(t-t')$. The velocity autocorrelation function decays exponentially as
   \begin{align}
	\langle v_0 \hat{\mathbf p}_i(t+T) \cdot v_0 \hat{\mathbf p}_i(t) \rangle &=   v_0^2 \, \mathrm e^{-D T} ,
 \end{align}
and the mean-squared displacement is given by:
  \begin{align}
 	\langle \left[ \mathbf r_i(t+T) - \mathbf r_i(t) \right]^2 \rangle &=  2 \frac{v_0^2}{D^2} \left( D T - 1 + \mathrm e^{-D T} \right) .
 \end{align}
The later expressions are used to fit the experimental data, Fig.~2E (main text) and Supplementary Figure~3 below, and provide the following values for the particle speed and diffusivity:
\begin{align}
\label{speed_mes}	&v_0=493\pm 17\,\rm \mu m.s^{-1} , \\
\label{diff}	&D^{-1}=0.31\pm 0.02\,\rm s .
\end{align}\\

\vspace{0.4cm}

\pagebreak

\noindent{\bf Supplementary Discussion 2 $\vert$ Population of interacting rollers}\\

\noindent {\bf Microscopic model. }
In \cite{Bricard2013}, we have theoretically described the microscopic dynamics of a population of colloids rolling on a conducting surface. We briefly summarize this model.  A colloid powered by the Quincke mechanism, when rotating close to a wall, exchanges momentum with this solid surface and translates at a speed given by
\begin{equation}
\label{speed}        v_0 = \frac{a \tilde \mu_t}{\mu_r \tau} \sqrt{\left( \frac{E_0}{E_{\rm Q}} \right)^2 -1} .
\end{equation}
$E_Q$ is the critical electric field below which the particle does not rotate. The Maxwell-Wagner time $\tau$ characterizes the dynamics of the electric charges at the colloid surface, which are responsible for the Quincke instability. $\mu_r$ and $\tilde \mu_t$ are mobility coefficients accounting for the viscous drag exerted by the liquid in the vicinity of solid wall that  depend logarithmically on the distance to the solid wall \cite{Bricard2013}.
In order to account for the roller-roller interactions at long distances, we have computed the electrostatic and hydrodynamic fields induced by the motion of a colloid. Assuming pairwise additive interactions, we have shown that these fields promote the alignement of the velocity of neighboring particles. Within this framework, the speed of the colloids is taken to be a constant  as it relaxes to $v_0$  much faster than the typical timescale of the orientation dynamics. In addition to these far-field couplings, we model the short-distance repulsion between colloids by an effective hard-core potential. The resulting equations of motion are given by Eqs.~[1]--[3] in the main text, where the functions $A(r)$, $B(r)$, $C(r)$ are given by:
\begin{align}
	A(r)&=A_1\left(\frac{a}{r}\right)^3\Theta(r)+A_2\left(\frac{a}{r}\right)^5\Theta(r),\\
	B(r)&=B_1\left(\frac{a}{r}\right)^4\Theta(r),\\
	C(r)&=C_1\left[2\left(\frac{a}{H}\right)\left(\frac{a}{r}\right)^2+\left(\frac{a}{r}\right)^3\Theta(r)\right] +C_2\left(\frac{a}{r}\right)^5\Theta(r) .
\end{align}
These electrostatic and hydrodynamic couplings are exponentially screened over a distance set by the channel height $H$ (see Fig.~1A, main text). For sake of simplicity, we approximate the screening function $\Theta(r)$ by the step function $\Theta(r) = 1$ if $r \leq H/\pi$ and $\Theta(r) = 0$ otherwise. In addition, the coefficients of the above functional forms are given by:
\begin{align}
	 &A_1 = 3 \tau^{-1} \tilde \mu_s ,\\
	 &A_2 = 9 \tau^{-1}  \left( \frac{\mu_\perp}{\mu_r} -1 \right) \left( \chi^\infty + \frac{1}{2} \right) \left( 1 - \frac{E_{\rm Q}^2}{E_0^2} \right) ,\\
        &B_1 =  6 \tau^{-1}  \left( \frac{\mu_\perp}{\mu_r} -1 \right) \sqrt{\frac{E_0^2}{E_{\rm Q}^2}-1} \left[\left( \chi^\infty + \frac{1}{2}\right) \frac{E_{\rm Q}^2}{E_0^2} - \chi^\infty\right]  ,\\
        &C_1 = A_1 ,\\
	&C_2 = \frac{5}{3} A_2.
\end{align}
$\mu_r$ and $\mu_\perp$ are mobility coefficients which only depend on the viscosity of the liquid and the gap $d$ between the colloid and the surface. $\chi^\infty$ depends on the dielectric permittivities $\epsilon_l$ and $\epsilon_p$ of the liquid and of the particles: $\chi^\infty = \left(\epsilon_p - \epsilon_l\right)\left(\epsilon_p+2\epsilon_l\right)$
The derivation of this model is provided in the supplementary informations of~\cite{Bricard2013}.\\

\noindent {\bf Estimation of the simulation parameters. }
In order to perform numerical simulations relevant to our experimental conditions, we have estimated the coefficients of the equations of motion as follows. The speed $v_0$ and the rotational diffusivity $D$ have been deduced from the single-particle dynamics, Eqs.~\eqref{speed_mes} and~\eqref{diff}. The threshold electric field $E_Q$ is measured experimentally as the critical value at which the colloids start moving. We use typical values for the dielectric permittivities of hexadecane ($\epsilon_l = 2.2 \epsilon_0$) and PMMA colloids ($\epsilon_p = 2.6 \epsilon_0$)~\cite{Pannacci} to evaluate $\chi^\infty = 0.06$. The mobility coefficients are estimated. We assume the distance between a particle and the surface to be $d \sim 50 \, \mathrm{nm}$. Although this parameter is not controlled precisely, it only yields small corrections to the mobility coefficients in the limit $d \ll a$, and weakly impacts the particle dynamics. Using the expressions derived in~\cite{Goldman1, Goldman2, O'Neil, Liu}, we find $\tilde \mu_s = 0.30$, $\mu_\perp/\mu_r = 1.6$ and $\tilde \mu_t/\mu_r = 8.7 \times 10^{-2}$. Finally, the Maxwell-Wagner time $\tau$ was calculated from Eq.~\eqref{speed}: $\tau = 0.29 \, \mathrm{ms}$. As a result, we obtain the following values for the microscopic coefficients: $A_1 = B_1 = C_1 = 0.9 \, \tau^{-1}$, $A_2 = 1.0 \, \tau^{-1}$, and $C_2 = 1.7 \, \tau^{-1}$.\\

\vspace{0.4cm}

\renewcommand{\refname}{{\large Supplementary References}}

\end{document}